\documentstyle[12pt]{article}                            
                            
\newlength{\dinwidth}                            
\newlength{\dinmargin}                            
\def\lapproxeq{\lower .7ex\hbox{$\;\stackrel{\textstyle <}{\sim}\;$}}
\def\gapproxeq{\lower .7ex\hbox{$\;\stackrel{\textstyle >}{\sim}\;$}}
\def\be{\begin{equation}}                            
\def\ee{\end{equation}}                            
\def\bea{\begin{eqnarray}}                            
\def\eea{\end{eqnarray}}

\makeatletter                                                       
\def\fmslash{\@ifnextchar[{\fmsl@sh}{\fmsl@sh[0mu]}}
\def\fmsl@sh[#1]#2{%
\mathchoice                                                       
{\@fmsl@sh\displaystyle{#1}{#2}}%
{\@fmsl@sh\textstyle{#1}{#2}}%
{\@fmsl@sh\scriptstyle{#1}{#2}}%
{\@fmsl@sh\scriptscriptstyle{#1}{#2}}}
\def\@fmsl@sh#1#2#3{\m@th\ooalign{$\hfil#1\mkern#2/\hfil$\crcr$#1#3$}}
\makeatother     
                            
\begin{document}                            
\titlepage                            
\begin{flushright}                            
DTP/98/60 \\
DESY 98-156 \\
hep-ph/9901420 \\                            
January 1999 \\                            
\end{flushright}                            
                            
\vspace*{2cm}                            
                            
\begin{center}                            

{\Large \bf $\Upsilon$ photoproduction at HERA compared to} \\          
          
\vspace*{0.7cm}          
{\Large \bf estimates of perturbative QCD}           
                            
\vspace*{1cm}                            
A.D.~Martin$^a$, M.G.~Ryskin$^{a,b}$ and T.~Teubner$^c$\\

\vspace*{0.5cm}                            
$^a$ Department of Physics, University of Durham, Durham, DH1 3LE. \\
$^b$ Petersburg Nuclear Physics Institute, Gatchina, St.~Petersburg, 188350, 
Russia.\\
$^c$ Deutsches Elektronen-Synchrotron DESY, D-22603 Hamburg, Germany.

\end{center}                            
                            
\vspace*{2cm}                            
                            
\begin{abstract}                          
We estimate the cross section for $\gamma p \rightarrow \Upsilon p$ by two           
independent methods.  First, by studying the corrections to the naive leading-order           
QCD formula and, second, by using parton-hadron duality.  The estimates are in good           
agreement with each other and with the recent measurements of the cross section at           
HERA.           
\end{abstract}                           
                   
\newpage                           
               
The elastic photoproduction of $\Upsilon$ mesons has recently been measured                
for the first time both by the ZEUS \cite{ZEUS} and H1 \cite{H1} collaborations.                 
The observed cross section $B \sigma \equiv \sigma (\gamma p \rightarrow \Upsilon        
p)~B                
(\Upsilon \rightarrow \mu^+ \mu^-)$ is found to be rather large               
\bea               
\label{eq:a1}               
B \sigma (\Upsilon) \; = \; 13.3 \: \pm \: 6.0_{\; - \; 2.3}^{\; + \; 2.7} \: {\rm pb} &        
{\rm at} & \langle W \rangle \; = \; 120~{\rm GeV} \quad\quad ({\rm ZEUS~\cite{ZEUS}}), \\
& & \nonumber \\               
\label{eq:a2}               
B \sigma (\Upsilon) \; = \; 16.0 \: \pm \: 7.5 \: \pm \: 4.0 \: {\rm pb} & {\rm at} &        
\langle W \rangle \; = \; 160~{\rm GeV} \quad\quad ({\rm H1\ (preliminary)~\cite{H1}})
\eea
where $\Upsilon$ stands for the sum over the $\Upsilon (1S), \Upsilon (2S)$ and               
$\Upsilon (3S)$ states and where $\langle W \rangle$ is the average               
photon-proton centre-of-mass energy.  In both presentations it was noted that the               
measurements are higher than the predictions of various QCD models by about two               
standard deviations.              
              
In this paper we estimate elastic $\Upsilon$ photoproduction from perturbative QCD               
by two independent methods.  Both calculations are found to give cross sections               
which agree well with the data, and so we conclude that there is no disagreement with               
QCD.  The first method is to study the main corrections to the naive QCD formula               
\cite{R} for heavy vector meson photoproduction.  The second method makes use of               
parton-hadron duality similar to \cite{MRT}, where $\rho$ electroproduction 
at HERA was described successfully.\\            
              
\noindent {\bf Method I}              
              
The first method is based on the leading-order expression for the photoproduction of a               
heavy vector meson of mass $M_V$ \cite{R}              
\be              
\label{eq:a3}              
\left . \frac{d \sigma}{dt} \: (\gamma p \rightarrow V p) \right |_{t = 0} \; = \;               
\frac{\alpha_S^2 \Gamma_{ee}^V}{3 \alpha M_V^5} \: 16 \pi^3 \left [xg \left (x,               
\frac{M_V^2}{4} \right ) \right ]^2              
\ee              
where $\Gamma_{ee}^V$ is the partial width of the $V \rightarrow ee$ decay,               
$\alpha_S$ is the QCD coupling, $\alpha = 1/137$ is the QED coupling and $g (x,               
\mu^2)$ is the gluon density measured at $x = M_V^2/W^2$ and the scale $\mu =     
m_Q \simeq M_V/2$.  We discuss below four types of corrections to this naive     
formula.\\
              
\noindent (a)~~{\bf Relativistic corrections}              
              
A non-relativistic wave function for the heavy vector meson was used to               
derive (\ref{eq:a3}), so first we quantify the size of the relativistic corrections.  For               
$J/\psi$ photoproduction these corrections have been the subject of debate               
\cite{FKS1,RRML,FKS2}.  Since the velocity of the charm quarks can be sizeable in               
the $J/\psi$ meson $\left ( \langle v^2 \rangle \simeq \frac{1}{4} \right )$ the               
kinematical correction coming from the quark propagators suppresses the $J/\psi$               
cross section by a factor of about 2 according to Ref.~\cite{RRML} or up to a factor of        
10 in \cite{FKS2}.  However it was shown in \cite{RRML} that to a good        
approximation this correction is nullified in the cross section formula (\ref{eq:a3})        
since it has been written in terms of the $J/\psi$ mass, instead of the current quark        
mass $(2 m_c < M_\psi)$ which should have been used in perturbative QCD.  So it        
turns out that there are practically no $\langle v^2 \rangle$ corrections to the leading        
order result (\ref{eq:a3}).  This result is consistent with a more detailed study by        
Hoodbhoy \cite{H}.  Strictly speaking to estimate the $\langle v^2 \rangle$      
corrections        
we should also take into account the               
contributions coming from more complicated components of the vector meson wave               
function which contain one or more gluons besides the $q\bar{q}$ pair.  These               
components were included in a self-consistent way by Hoodbhoy \cite{H}, who               
showed that the total $O (v^2)$ relativistic correction (including corrections to the               
quark propagator and extra gluons in the $J/\psi$ wave function) to the naive cross               
section formula (\ref{eq:a3}) amount to at most only 7\% for $J/\psi$               
photoproduction.  The correction should be even smaller for $\Upsilon$ production.
Therefore we neglect these relativistic effects below. \\      
              
\noindent (b)~~{\bf Real part}              
              
Only the imaginary part of the $\gamma p \rightarrow V p$ amplitude is expressed in               
terms of $xg (x, \mu^2)$ in (\ref{eq:a3}).  To restore the real part we may use the               
approximation              
\be              
\label{eq:a4}              
{\rm Re} \: A/ {\rm Im} \: A \; \simeq \; \pi \lambda/2              
\ee              
where              
\be              
\label{eq:a5}              
\lambda \; = \; \frac{\partial \log A}{\partial \log s} \; = \; \frac{1}{xg (x, \mu^2)} \:               
\frac{\partial (xg (x, \mu^2))}{\partial \ln (1/x)}.              
\ee              
This approximation is valid for small $\lambda$, say $\lambda < \frac{1}{2}$, for the               
even signature amplitude $A$.  The real part correction factor              
\be              
\label{eq:a6}              
C_b \; = \; \left ( 1 \: + \: \frac{\pi^2 \lambda^2}{4} \right ) \; = \; 1.43 (1.36) \;               
{\rm or} \; 1.54 (1.46)              
\ee              
for $\langle W \rangle = 120 (160)$~GeV and scale $\mu^2 = 25$~GeV$^2$,  
according to whether the MRS(R2) \cite{MRS} or the GRV \cite{GRV} gluon distribution is  
used. \\      
              
\noindent (c)~~{\bf The effect of off-diagonal partons}              
              
At leading order the elastic photoproduction of vector mesons is mediated by       
two-gluon exchange.  Strictly speaking the amplitude is proportional to the off-      
diagonal               
gluon distribution $x^\prime g (x, x^\prime)$, where the momentum fraction               
$x^\prime$ carried by the \lq\lq second" gluon is much smaller than the fraction $x               
\simeq M_V^2/W^2$ carried by the \lq\lq first" gluon.  The effect, $R =               
x^\prime g (x, x^\prime)/xg (x)$, was estimated in Ref.~\cite{MR}.  The correction is               
almost negligible for $\rho$ production, giving about 10\% enhancement of the $J/\psi$ 
amplitude, but is much more important for $\Upsilon$ photoproduction since the scale               
$\mu^2 \simeq M_\Upsilon^2$ and the value of $x = M_\Upsilon^2/W^2$               
are much larger.  In fact the correction factor to the cross section formula (\ref{eq:a3})       
for $\Upsilon$ production is \cite{MR}       
\be              
\label{eq:a7}              
C_c \; = \; R^2 \: \simeq \: (1.4)^2 \: \simeq \: 2.              
\ee              
              
\noindent (d)~~{\bf NLO corrections}              
              
Finally we have NLO corrections which come from an explicit integration over the               
loop which corresponds to the convolution of the exchanged gluons with the vector               
meson wave function.  The integral is not of a pure logarithmic form, which was       
assumed in deriving the leading order result (\ref{eq:a3}).  The typical range of       
integration corresponds to the scale $\mu^2 = \frac{1}{4} M_V^2$ and the               
non-logarithmic contribution is well approximated by a NLO correction with               
a coefficient of about $\frac{1}{2}$ \cite{RRML}.  That is the correction can be written 
as
\be              
\label{eq:a8}              
C_d \; \simeq \; \left (1 \: + \: \alpha_S \left (\textstyle{\frac{1}{4}} M_V^2   
\right ) /2 \right ) \;               
\simeq \; 1.2              
\ee              
for $\Upsilon$ photoproduction.              
              
We are now in the position to estimate the cross section for elastic $\Upsilon$               
photoproduction.  If we assume that the elastic $\Upsilon$ differential cross section,       
$d \sigma/dt \sim \exp (bt)$, has a slope $b = 4$~GeV$^{-2}$, as for elastic $J/\psi$               
photoproduction, then the leading order formula gives              
\be              
\label{eq:a9}              
\left . B \sigma (\Upsilon_{1S})  \right |_{\rm LO} \; \simeq \; 1.7 \: {\rm pb}              
\ee              
at $\langle W \rangle = 120$~GeV.  Taking into account the corrections               
$C_{b,c,d}$ we obtain              
\be              
\label{eq:a10}              
B \sigma (\Upsilon_{1S}) \; \simeq \; 6 \: {\rm pb}.              
\ee              
However the cross sections measured by the ZEUS and H1 collaborations,               
(\ref{eq:a1}) and (\ref{eq:a2}), did not select the pure $\Upsilon (1S)$ state, but rather               
the data were integrated over an interval of the $\mu^+ \mu^-$ mass which includes at               
least the $1S$, $2S$ and $3S$ resonances.  We estimate the combined contributions of               
the $2S$ and $3S$ states to be 40\% of the $\Upsilon (1S)$ cross section, in               
agreement with the measurement of the CDF collaboration \cite{CDF}.  Thus               
multiplying (\ref{eq:a10}) by a factor of 1.4 we estimate              
\be              
\label{eq:a11}              
B \sigma (\Upsilon) \; = \; 8.4 \: {\rm pb},              
\ee              
which is compatible with the ZEUS measurement given in (\ref{eq:a1}).  Going from               
$\langle W \rangle = 120$~GeV to 160~GeV, the cross section is predicted to               
increase by a factor of 1.6, due to the larger gluon density at smaller $x$, and so the               
prediction is in agreement with the preliminary H1 measurement quoted in       
(\ref{eq:a2}).  \\            
              
\noindent {\bf Method II}              
              
The second method that we use to estimate the elastic photoproduction of vector               
mesons from perturbative QCD is based on parton-hadron duality.  The procedure               
\cite{MRT} is to calculate the amplitude for open $q\bar{q}$ production, then to               
project the amplitude onto the $J^P = 1^-$ $q\bar{q}$ state and finally to integrate               
the cross section over an appropriate interval $\Delta M$ of the mass of the     
$q\bar{q}$ pair which               
includes the resonance peak.  As there are almost no other possibilities for               
hadronization at $M_{q\bar{q}} \simeq M_V$ we should obtain a reasonable               
estimate of the cross section for vector meson production.  This framework was found               
\cite{MRT} to describe successfully the energy and $Q^2$ dependence of               
$\rho$-meson electroproduction $(\gamma^* p \rightarrow \rho p)$, both for               
longitudinally and transversely polarized $\rho$ mesons, including the $Q^2$               
dependence of the $\sigma_L/\sigma_T$ ratio.              
              
The computer code used here to estimate the $\gamma p \rightarrow 
\Upsilon p$ cross section takes into account all the corrections 
mentioned above, with the exception of the enhancement due to the 
use of off-diagonal gluons.  (The      
enhancement is negligible for $\rho$ production.)~~We therefore include the extra      
factor $C_c \simeq 2$ of (\ref{eq:a7}).  As the recent measurements (\ref{eq:a1}) and      
(\ref{eq:a2}) cover a rather large $\mu^+ \mu^-$ mass interval\footnote{Note that in the      
parton-hadron duality approach $\Delta M$ is related to the formation time rather      
than to the width of the resonances.  The $\gamma p \rightarrow \Upsilon p$ data are      
integrated over an interval larger than the formation time but which includes at least      
the first three $\Upsilon$ states and possibly some small contribution from the      
$b\bar{b}$ continuum.} ($\Delta M \sim 2$~GeV) over the $\Upsilon$ resonances a      
description based on parton-hadron duality might even be more appropriate to      
describe the data.              
    
In practice we have to extend our code to include the mass $m$ of the heavy $b$  
quark.  First we restore the mass term $m \delta_{\lambda\lambda^\prime}$ in the  
expression     
for the $\gamma \rightarrow q\bar{q}$ matrix element, see Eq.~(32) of     
Ref.~\cite{LMRT}.  Here we will use the notation $\lambda = i = +, -$ and     
$\lambda^\prime = i^\prime = +, -$ for the helicities of the quark and antiquark.  The  
cross section is then given by (see also \cite{LMRT}) 
\be 
\label{eq:x11} 
\left . \frac{d^2 \sigma^T}{dM^2 dt} \right |_{t = 0} \; = \; \frac{\pi^2 e_q^2  
\alpha}{3 (Q^2 + M^2)^2} \: \int \: 2 dz \: \sum_{i, i^\prime} \: |B_{i i^\prime}|^2 
\ee 
where the helicity amplitudes are (for photon helicity $+1$) 
\be 
\label{eq:y11} 
B_{+ +} \; = \; \frac{m I_L}{2 \sqrt{z (1 - z)}}, \quad\quad B_{- -} \; = \; 0 
\ee 
and 
\be 
\label{eq:z11} 
B_{+ -} \; = \; \frac{-z k_T I_T}{\sqrt{z (1 - z)}}, \quad\quad B_{- +} \; = \; \frac{(1 -  
z) k_T I_T}{\sqrt{z (1 - z)}}. 
\ee 
The variable $z$ is the momentum fraction carried by the quark and $k_T$ is its  
transverse momentum.  The integrals $I_L$, $I_T$ are defined in \cite{LMRT}.  The  
second modification due to the mass of the quark is associated with the fact that the  
$\gamma \rightarrow q\bar{q}$ helicity amplitudes $B_{ii^\prime}$ are defined in  
the proton  
rest frame (pRF), while the projection onto the $J^P = 1^-$ state was done using the  
quark helicities in the $q\bar{q}$ rest frame ($q\bar{q}$RF).  Unfortunately helicity  
is not a good quantum number for a heavy quark.  It can be changed by a Lorentz  
boost.  So we have to compute the helicity amplitudes $A_{jj^\prime}$ in the  
$q\bar{q}$RF in terms of $B_{ii^\prime}$ in the pRF.  We have    
\be    
\label{eq:b11}    
A_{jj^\prime} \; = \; \sum_{i,i^\prime} \: c_{ij} \: c_{j^\prime i^\prime} \:     
B_{ii^\prime}    
\ee    
where we will calculate the coefficients ($c_{ij}$ for the quark 
and $c_{j^\prime i^\prime}$ for the antiquark) via the polarized quark density     
matrix $\rho$.  For a quark with 4-momentum $k_\mu$ and polarisation vector     
$a_\mu$ (satisfying $a^2 = -1$ and $a.k = 0$) we have \cite{IZ}    
\be    
\label{eq:c11}    
\rho \; = \; \frac{(\fmslash{k} + m)}{2} \: \frac{(1 + \gamma_5 \fmslash{a})}{2},    
\ee    
where $(1 + \gamma_5 \fmslash{a})/2$ projects onto the state with polarisation vector     
$a_\mu$.  For the states with helicities $j = \pm$ in the $q\bar{q}$RF these vectors     
take the form    
\be    
\label{eq:d11}    
a_\mu^j \; = \; (a_0; \mbox{\boldmath $a$}_T, a_z) \; = \; (a_0; \mbox{\boldmath     
$a$}) \; = \; \pm (k; k_0 \mbox{\boldmath $k$}/k)/m    
\ee    
where $k = | \mbox{\boldmath $k$}|$.  Let $b_\mu^i$ be the analogous polarisation     
vectors describing helicities $i = \pm$ in the pRF.  After the boost from the pRF to the     
$q\bar{q}$RF these vectors are given by    
\be    
\label{eq:e11}    
b_\mu^i \; = \; \pm \left ( \frac{M}{2m} \: - \: \frac{m}{zM}; \:     
\frac{\mbox{\boldmath $k$}_T}{m}, \: \frac{M}{m} \left (z - \textstyle{\frac{1}{2}}   
\right ) \: + \: \frac{m}{zM} \right )    
\ee    
where $M$ is the mass of the $q\bar{q}$ pair, $z = \frac{1}{2} + k \cos \theta/M$  
and $\theta$ is the quark decay angle in the $q\bar{q}$RF.    
    
Now that we have all the polarisation vectors in the $q\bar{q}$RF we can calculate     
the product of the two different projectors\footnote{We omit the projector     
$\frac{1}{2} (\fmslash{k} + m)$ of (\ref{eq:c11}) so that the trace includes both     
quark and antiquark components.  Hence the factor 2 in $2 c_{ij}^2$ on 
the right hand side of~(\ref{eq:f11}).}    
\be    
\label{eq:f11}    
\textstyle{\frac{1}{4}} {\rm Tr} \left [ (1 + \gamma_5 \fmslash{a}^j) (1 + \gamma_5     
\fmslash{b}^i) \right ] \; = \; 1 - (a . b) \; = \; 2 c_{ij}^2.    
\ee    
Thus we have    
\be    
\label{eq:g11}    
c_{+ +} \; = \; c_{- -} \; = \; c_{+ -} \; = \; -c_{- +} \; = \; \sqrt{\frac{1 - (a .     
b)}{2}},    
\ee    
where $c_{- +}$ is chosen to be negative to ensure the orthogonality of the $b^i$     
states when expressed in the $a^j$ basis.  Note that if the quark mass $m \rightarrow     
0$ then     
\be    
\label{eq:h11}    
a_\mu^\pm \; \rightarrow \; b_\mu^\pm \; = \; \pm \: k_\mu/m    
\ee    
leading to the unit matrix $c_{ij} = \delta_{ij}$.    
    
The procedure is therefore to decompose the original amplitude $B_{ii^\prime}$ in     
the pRF in terms of the 
amplitudes $A_{jj^\prime}$ with helicities $j, j^\prime = +, -$ in     
the $q\bar{q}$RF, as in (\ref{eq:b11}).  The resulting helicity amplitudes are  
projected onto the $J_m^P = 1_m^-$ $q\bar{q}$ states so as to obtain the production  
amplitudes in spin states $|1, m \rangle$ with $m = 0, \pm 1$    
\be    
\label{eq:i11}    
T_m \; = \; \sum_{j, j^\prime} \: \sqrt{\frac{3}{2}} \: \int_{-1}^1 \: d \cos \theta \:  
A_{jj^\prime} \: d_{1m}^1 (\theta) \: \delta_{m, (j - j^\prime)}. 
\ee    
The cross section of diffractive $J^P = 1^-$ $q\bar{q}$ pair production is then  
obtained by summing over the $m = 0, \pm 1$ amplitudes squared in an analogous  
way to that described in Ref.~\cite{MRT}. 
 
The expressions for $I_{L,T}$ of (\ref{eq:y11}) and (\ref{eq:z11}) were written in  
\cite{LMRT} for the (dominant) imaginary part of the amplitude.  They are integrals  
over the gluon transverse momentum $\ell_T$ with weight $w$ corresponding to the  
$q\bar{q}$-loop 
\be 
\label{eq:j11} 
{\rm Im} I_{L,T} \; = \; \int \: d\ell_T^2 \: f (x, \ell_T^2) \: w_{L,T} (\ell_T^2, Q^2, 
M^2  
\cdots ) 
\ee 
where $f$ is the unintegrated gluon distribution 
\be 
\label{eq:k11} 
f (x, \ell_T^2) \; = \; \frac{\partial xg (x, \ell_T^2)}{\partial \ell_T^2}. 
\ee 
The real part of the amplitude has been included using the same approximation as in  
Eqs.~(\ref{eq:a4}, \ref{eq:a5}), by 
replacing $f (x, \ell_T^2)$ with the derivative $(\pi/2)  
\partial f (x, \ell_T^2)/\partial \ln (1/x)$, that is we have computed 
\be 
\label{eq:l11} 
{\rm Re} I_{L,T} \; = \; \frac{\pi}{2} \: \int \: d \ell_T^2 \: \frac{\partial f (x,  
\ell_T^2)}{\partial \ln 1/x} \: w_{L,T} (\ell_T^2, Q^2, M^2, \ldots ). 
\ee 
Finally the NLO corrections (analogous to $C_d$ of (\ref{eq:a8})) are approximated in  
the code by use of a $K$-factor, see \cite{MRT,LMRT}.  As 
in \cite{LMRT} (and (\ref{eq:a8})) the 
scale in the coupling $\alpha_S (\mu^2)$ was chosen to be $\mu^2 = M^2/4$. 
              
To compare with the ZEUS data for $\gamma p \rightarrow \Upsilon p$ we integrate               
over the mass interval 8.9--10.9~GeV.  We take the mass of the $b$ quark to be $m_b               
= 4.6$~GeV and, as in Ref.~\cite{MRT}, we use the MRS(R2) set of partons               
\cite{MRS}.  In this way we find              
\be              
\label{eq:a12}              
\sigma (\gamma p \rightarrow \sum \Upsilon p) \; \simeq \; 620 \: {\rm pb}              
\ee              
at $\langle W \rangle = 120$~GeV.  This should be compared with the value $635 \pm               
310$~pb obtained from the ZEUS measurement.  The \lq\lq ZEUS\rq\rq~value is     
calculated from               
(\ref{eq:a1}) by assuming the ratios     
\be              
\label{eq:a13}              
B \sigma (\Upsilon_{1S}) \: : \: B \sigma (\Upsilon_{2S}) \: : \: B \sigma      
(\Upsilon_{3S}) \; \simeq \; 0.7 \: : \: 0.15 \: : \: 0.15,              
\ee              
which are consistent with the expectations of (\ref{eq:a3}) and with the CDF data               
\cite{CDF}, and using the known leptonic branching ratios of the $\Upsilon$ states.                
What is the uncertainty in our cross section estimate (\ref{eq:a12}) due to the choice      
of $m_b$?  The PDG \cite{PDG} give $4.1 < m_b < 4.4$~GeV for the running mass      
evaluated at $\mu = m_b$ in the $\overline{\rm MS}$ scheme, which corresponds to      
the range $4.5 < m_b < 4.8$~GeV for the pole mass.  The latter mass is relevant for      
our perturbative calculations.  If we were to use $m_b = 4.4$ or 4.8~GeV then the      
cross section (\ref{eq:a12}) would become 900 or 380 pb respectively.             
           
Fig.~1 shows the ZEUS and H1 data, together with our QCD estimates obtained            
from the two independent methods.  The comparison is given for the $\gamma p            
\rightarrow \Upsilon (1S) p$ cross section as a function of the photon-proton            
centre-of-mass energy $W$.  To obtain the $\Upsilon (1S)$ cross section for method            
II, in which we integrate over the mass interval 8.9--10.9~GeV, we divide the            
$\gamma p \rightarrow (\sum \Upsilon) p$ result by a factor 1.7.  This factor is            
obtained using (\ref{eq:a13}).  For completeness we also show in Fig.~1 the            
prediction from the naive leading order formula (\ref{eq:a3}) of 
method I as well as the result of method II without the corrections 
from the real part of the amplitudes.  We note that the            
$\gamma p \rightarrow \Upsilon p$ process is particularly sensitive to off-diagonal gluon      
effects.  These effects give an enhancement of about a factor of 2, which appears to be      
required by the data.           
     
To estimate the ratio of $\Upsilon$ to $J/\psi$ photoproduction we calculate both 
cross sections using all the corrections listed for method I.  For $\langle W      
\rangle = 120$~GeV we find 
\be     
\label{eq:a14}     
\sigma (\Upsilon_{1S})/\sigma (J/\psi) \; \simeq \; 1/400     
\ee     
where the MRS(R2) gluon has been used.  This is near the lower bound observed by ZEUS      
\cite{ZEUS}.  To be specific we underestimate $\Upsilon$ production and      
overestimate\footnote{Note that in Ref.~\cite{RRML} absorptive corrections were      
included and that a better description of $J/\psi$ photoproduction was obtained.  Here      
absorptive corrections were not discussed as they should be small for $\Upsilon$      
photoproduction.} $J/\psi$ production.  We note that if, for example, the steeper GRV      
gluon was used the ratio would be 1/1500.  Clearly, as expected, the ratio is very      
sensitive to the $x$ and scale dependence of the gluon in the region of very small $x$      
and low scale that is sampled by $J/\psi$ production.  In the case of $\Upsilon$     
production we sample $x \sim 0.01$ and $Q^2 \sim 25$~GeV$^2$ where the     
predictions are much more stable with respect to different parametrizations of the     
gluon.     
           
We conclude that the recent measurements of elastic $\Upsilon$ photoproduction are            
in good agreement with the expectations of perturbative QCD.  Of course when more            
precise data become available the effects that are estimated here --- relativistic      
corrections, NLO contributions, off-diagonal parton effects and the contribution of the      
real part of the amplitude --- should be calculated in detail.  The elastic      
photoproduction of $J/\psi$ and $\Upsilon$ at HERA will be a good laboratory to      
probe these effects. \\     
     
\noindent {\bf Acknowledgements}     
     
M.G.R. thanks the Royal Society, INTAS (95--311) and the Russian Fund of      
Fundamental Research (98~02~17629) for support. \\

\noindent {\bf Note added}     
     
While writing this note we received a paper on the same 
subject by Frankfurt, McDermott and Strikman~\cite{FMS}.  Their 
procedure is comparable to our method I and also yields a cross 
section in agreement with the HERA data.  However although our 
final values are similar, there are large, physically significant, 
differences in the various component factors.  Unlike the present 
work, they do not make use of the result of Hoodbhoy~\cite{H} and 
so have a large suppression from the relativistic corrections.  
In \cite{FMS} this suppression is compensated by a larger off-diagonal 
effect (2.6 as compared to 2), a larger real part and an enhancement 
due to re-scaling (giving an effective $Q^2$ of about 40 as compared 
to 25 GeV$^2$).

\vspace{1cm}
{\large\bf
\noindent Figure Caption}           
           
\begin{itemize}           
\item[Fig.~1] The $\gamma p \rightarrow \Upsilon (1S) p$ cross 
section as a function of the $\gamma p$ centre-of-mass 
energy $W$.  The data points are the ZEUS \cite{ZEUS} and 
preliminary H1 \cite{H1} measurements.  The horizontal 
lines indicate the range of $W$ sampled by the measurements, 
the vertical lines the systematic and statistical errors added 
linearly.  The continuous curves are obtained from the two QCD 
calculations (model I and II) described in the text.  The 
dotted curve corresponds to the naive leading-order formula 
(\ref{eq:a3}), the dashed curve shows our prediction for 
method II if we neglect the corrections from the real part 
of the amplitudes.
\end{itemize}            

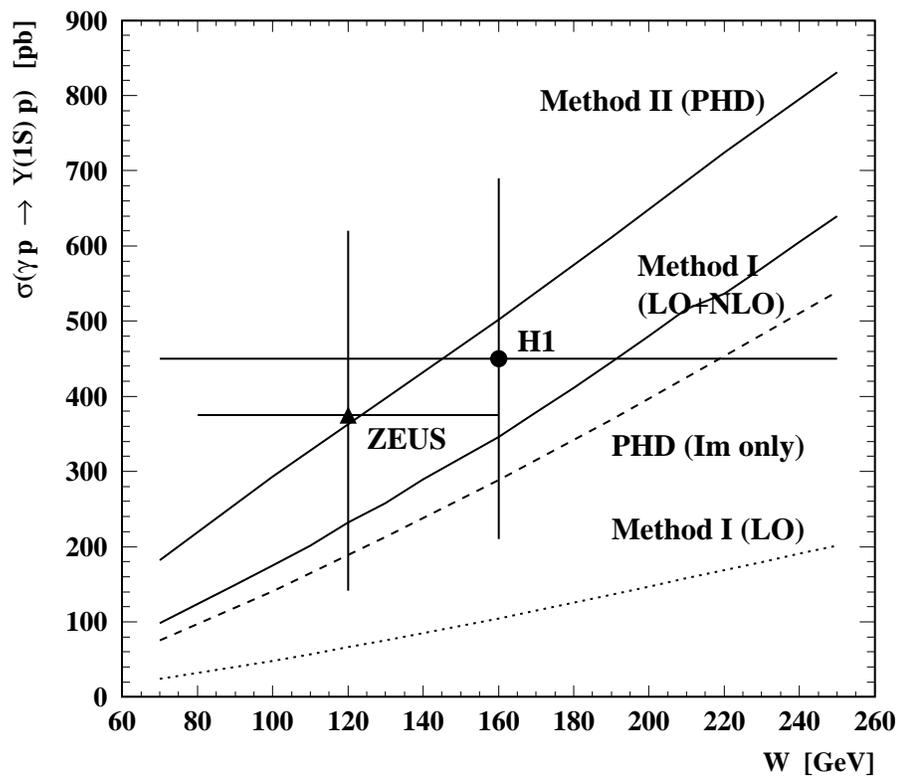
\end{document}